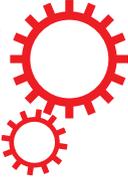

OPEN

# *In situ* Characterization of Nanoparticles Using Rayleigh Scattering

Biswajit Santra[1,*], Mikhail N. Shneider[2,*] & Roberto Car[1,3,4,*]

We report a theoretical analysis showing that Rayleigh scattering could be used to monitor the growth of nanoparticles under arc discharge conditions. We compute the Rayleigh scattering cross sections of the nanoparticles by combining light scattering theory for gas-particle mixtures with calculations of the dynamic electronic polarizability of the nanoparticles. We find that the resolution of the Rayleigh scattering probe is adequate to detect nanoparticles as small as $C_{60}$ at the expected concentrations of synthesis conditions in the arc periphery. Larger asymmetric nanoparticles would yield brighter signals, making possible to follow the evolution of the growing nanoparticle population from the evolution of the scattered intensity. Observable spectral features include characteristic resonant behaviour, shape-dependent depolarization ratio, and mass-dependent line shape. Direct observation of nanoparticles in the early stages of growth with unobtrusive laser probes should give insight on the particle formation mechanisms and may lead to better-controlled synthesis protocols.

Since their discovery in 1991[1] carbon nanotubes have attracted great attention due to the expectation that they could lead to electronic devices with unprecedented capabilities. So far this expectation has not been fulfilled, mainly because it has not been possible to access large yields of single-walled nanotubes (SWNT) with a well-defined chirality[2–7], which is the key factor governing their electronic properties[8]. Defect-free SWNTs are typically extracted from synthesis products that contain a mixture of defective and non defective nanotubes[9–12]. Thus one can collect nanotubes with desired properties only on a small scale and at a high cost. A possible way out of this difficulty requires controlling the chirality directly in the synthesis process. Some progress in this direction was made recently by employing techniques such as catalytic control[2–4], precursor selection[5], vapor phase epitaxy[6], or nanoring templates[7] in the context of chemical vapor deposition (CVD) synthesis.

A major stumbling block on the way of devising controlled synthesis processes is the lack of detailed microscopic knowledge of the processes responsible for the growth of the carbon nanoparticles[13]. To better understand these processes experimental techniques that can characterize the particles *in situ* during the early stages of growth would be extremely useful. So far *in situ* techniques have been limited to a few cases. In one example, time-resolved images of nanotubes nucleating on metal catalysts were taken by environmental transmission electron microscopy (TEM)[14,15]. In another case, the growth rates of nanotubes were measured by Raman spectroscopy[16,17]. These examples refer to CVD growth. We are not aware of any example in the context of arc synthesis, which takes place at substantially higher operating temperatures and, partly because of that, produces relatively higher yields of structurally pristine SWNTs[18–24].

Laser-based optical imaging, such as Rayleigh, Raman, and photoluminescence, possess advantages over TEM because they are non-intrusive, allow for high throughput imaging, and provide spectroscopic information on electronic and vibrational properties[25]. So far, optical properties of SWNTs have been measured after synthesis, i.e. *ex situ*, by Raman[26–28], photoluminescence[28–30], Rayleigh[31–37], and absorption spectroscopy[38–41]. Rayleigh scattering (RS) spectroscopy probes the electronic polarizability of the scattering particles. Since RS relies on elastic scattering its signal is several orders of magnitude brighter than that of inelastic scattering techniques, such as Raman and photoluminescence spectroscopies. RS spectroscopy can be applied to both semiconducting and metallic nanotubes[32–34], and can be used to excite and detect radiation in the visible range for better spatial resolution and detection sensitivity[35]. Indeed, a previous study[42] by one of us concluded that a relatively small concentration of $C_{60}$ in Argon (Ar) atmosphere at ambient conditions would be detectable by RS spectroscopy. This work used a static estimate for the polarizability of $C_{60}$. Here we extend this analysis to arc discharge conditions, use the full dynamic polarizability of the nanoparticles, and consider the effects of the size and shape of the nanoparticles as well.

[1]Department of Chemistry, Princeton University, Princeton, NJ 08544, USA. [2]Mechanical and Aerospace Engineering Department, Princeton University, Princeton, NJ 08544, USA. [3]Princeton Institute for the Science and Technology of Materials, Princeton University, Princeton, NJ 08544, USA. [4]Program in Applied and Computational Mathematics, Princeton University, Princeton, NJ 08544, USA. *These authors contributed equally to this work. Correspondence and requests for materials should be addressed to R.C. (email: rcar@princeton.edu)

In the early stages of nanotube growth, in arc discharge or plasma assisted CVD synthesis, the chamber contains a mixture of nanoparticles and gas filling molecules. The typical filling gas that we consider here is helium (He), but other inert gases, like e.g. Ar, could also be considered. In the early stages of growth the nanoparticles are expected to be small in size (few nm), asymmetric, and have small concentration relative to the filling gas. These conditions are suitable for RS, which requires that the size of the particles be much smaller than the wavelength of the incident light. The corresponding scattering intensity is controlled by the concentration and the electronic polarizability of the different particles and molecules that constitute the gas-particle mixture present in the synthesis region of the chamber[43–45]. He molecules are monoatomic and their polarizability is isotropic, essentially constant and equal to its static limit in the frequency range corresponding to typical laser probes. By contrast, the nanoparticles have an anisotropic polarizability tensor that reflects their asymmetric shapes and may have non-negligible frequency dependence in the spectral window of interest. The anisotropy of the polarizability induces a measurable depolarization of the scattered light[46]. Moreover, the particles are multiatomic and their polarizability should be substantially larger than that of the He atoms, facilitating the detection of the particle contribution to the RS signal over the background.

In order to find out whether RS could be used to monitor *in situ* the growth of nanoparticles we compute the relevant cross sections by combining electronic structure and dynamical light scattering theory for gas-particle mixtures, at the expected thermodynamic conditions in the arc periphery region where particle formation and growth takes place. We limit ourselves to carbon nanoparticles and consider a few selected single shell nanoparticles with different aspect ratios. We compute their frequency-dependent electronic polarizability using time-dependent density functional theory (TDDFT), a scheme that yields the time evolution of the spatial density of a many-electron system subject to an external time varying field[47]. While in principle TDDFT is an exact approach, in practice its accuracy depends on the approximation adopted for the exchange-correlation functional[48–54]. Published results on the optical absorption spectra of fullerenes[55–57] and of carbon nanotubes[53,54,58] suggest that standard functional approximations should be sufficient for our purpose. Our own calculations of the RS spectra of selected carbon nanotubes for which experimental information is available further support this proposition.

Here we focus on the RS cross section of carbon nanoparticles of finite length, finding that the longitudinal polarizability of nanoparticles with different aspect ratios has resonances at frequencies that depend on the size and shape of the nanoparticle, originating characteristic signatures in the RS intensity and depolarization ratio. We also find that these features should be clearly distinguishable from the background signal for particles as small as $C_{60}$ at concentrations in excess of 1 particle per million of gas molecules. These concentrations are in excess of estimates based on the measured carbon deposits on a substrate placed in the arc periphery[59]. Nanoparticles more asymmetric than those considered in our study should be detectable at even lower concentrations. Moreover, at the elevated temperatures of the arc chamber Doppler broadening results in line shapes that are strongly mass dependent. Thus it should be possible to monitor the time evolution of the mass of the growing nanoparticles from the line shape of the RS signal.

The detectability, demonstrated here, of the RS signals of nanoparticles under arc discharge conditions of synthesis, combined with the sensitivity of the spectral intensity on the shape, size, concentration, mass, and temperature of the growing nanoparticles suggests that RS spectroscopy should be an ideal tool for *in situ* characterization during synthesis. The experiments that we advocate should contribute to better elucidate the mechanisms that lead to the formation and growth of nanoparticles under arc discharge. Such understanding is a prerequisite for better controlling the synthesis process.

## Results and Discussion

**Rayleigh Scattering.** Rayleigh scattering is caused by a polarizable particle of size much smaller than the wavelength of light. The oscillating electric field of the incident light induces an oscillating dipole moment in the particle, which radiates light in dipolar pattern. For a spherically symmetric molecule and linearly-polarized light, the induced dipole moment is parallel to the incident polarization direction and is proportional to the polarizability of the particle. If the molecule is not spherically symmetric the induced dipole moment depends on the orientation of the molecule and the scattered light is partially depolarized. In Fig. 1 a schematic diagram shows the scattering of a vertically polarized incident light beam by a gas-particle mixture. As the particles can rotate freely, the scattering intensity is averaged over all the orientations of the particles with respect to the incident electric field. For a vertically polarized incident laser beam, the scattered intensity in the direction $\theta = \pi/2$ and $\varphi = \pi/2$, is polarized either vertically ($I_{VV,n}$) or horizontally ($I_{VH,n}$). $I_{VV,n}$ and $I_{VH,n}$ are given by[43]:

$$I_{VV,n}(\lambda) = \frac{n_n V \pi^2 I_L}{\varepsilon_0^2 \lambda^4 r^2} \left( \frac{45 \alpha^2(\lambda) + 4\gamma^2(\lambda)}{45} \right), \tag{1}$$

and

$$I_{VH,n}(\lambda) = \frac{n_n V \pi^2 I_L}{\varepsilon_0^2 \lambda^4 r^2} \left( \frac{\gamma^2(\lambda)}{15} \right), \tag{2}$$

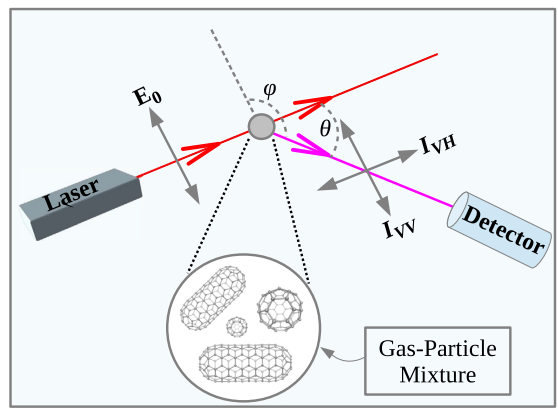

**Figure 1.** Schematic diagram of a Rayleigh scattering experiment. A vertically polarized incident laser beam is scattered from a gas-particle mixture. $I_{VH}$ and $I_{VV}$ denote the intensity of the horizontally and vertically polarized scattered light, respectively. $\theta$ is the scattering angle. $\varphi$ is the angle between the polarization vector of the incident light ($\mathbf{E}_0$) and the propagation direction of the scattered light.

respectively, where $n_n$ is the concentration of the nanoparticles, $V$ is the scattering volume, $r$ is the distance of the detector from the scatterer, $\varepsilon_0$ is the dielectric permittivity of the vacuum, $I_L$ and $\lambda$ are the intensity and the wavelength of the incident laser radiation. $\alpha(\lambda)$ and $\gamma(\lambda)$, i.e. respectively the average and anisotropic components of the frequency-dependent polarizability tensor, are defined as:

$$\alpha(\lambda) = \frac{1}{3}(2\alpha_\perp(\lambda) + \alpha_\parallel(\lambda)) \text{ and } \gamma(\lambda) = \alpha_\parallel(\lambda) - \alpha_\perp(\lambda). \tag{3}$$

Here $\alpha_\parallel$ is the longitudinal polarizability along the long axis of a nanoparticle and $\alpha_\perp$ is the corresponding transverse polarizability. The intensity of the total scattered radiation by the nanoparticles, $I_{tot,n}(\lambda)$, is the sum of the intensities of the horizontally and vertically polarized scattered light, i.e., $I_{tot,n}(\lambda) = I_{VH,n}(\lambda) + I_{VV,n}(\lambda)$. Thus, in the direction $\theta = \pi/2$ the total scattering intensity is

$$I_{tot,n}(\lambda) = \frac{n_n V \pi^2 I_L}{\varepsilon_0^2 \lambda^4 r^2} \left( \frac{45\alpha^2(\lambda) + 7\gamma^2(\lambda)}{45} \right). \tag{4}$$

The RS intensity depends on the frequency-dependent polarizability of the nanoparticles, which is computed using TDDFT (see Methods).

**Accuracy of TDDFT Calculations.** To estimate quantitatively the frequency-dependent polarizability of carbon nanoparticles ranging from fullerenes to nanotubes of various lengths we use TDDFT, an approach that is well suited for this task as it is non-empirical and has proven predictive capability[48–58]. The accuracy of TDDFT calculations depends on the approximation followed for the exchange and correlation kernel. Typically one adopts the adiabatic approximation, which allows one to use static exchange-correlation functionals for dynamic calculations. Commonly used static functionals are based on the generalized gradient approximation (GGA) or on the hybrid functional approximation[50–52]. In the case of fullerenes, TDDFT calculations typically underestimate the experimental optical absorption energies by ~0.3 eV and yield associated oscillator strengths that are within 25% of experiment[55,57]. TDDFT absorption spectra of infinite periodic nanotubes are also in qualitatively good agreement with experiment[58]. The GGA absorption energies of nanotubes are affected by errors similar to those observed in the fullerenes, while hybrid functionals reduce the deviations from experiment to ~0.1 eV or less[53,54].

We are not aware of experimental RS data for fullerenes but RS experiments on selected isolated SWNTs of specific chirality have been reported in the literature[32–34]. It is of interest to further assess the accuracy of our methodology in these cases. We focus here in particular on the semiconducting (15,10) and the metallic (10,10) SWNTs. In the experiments fully-grown SWNTs were suspended on a fabricated surface and illuminated one at a time with a focused laser beam with polarization parallel to the axis of the nanotube[33]. For long SWNTs the longitudinal polarizability ($\alpha_\parallel$) is much larger than the transverse polarizability ($\alpha_\perp$), hence the total RS intensity is proportional to $\alpha_\parallel^2(\lambda)/\lambda^4$. Calculated RS spectra with the Perdew-Burke-Ernzerhof (PBE[60]) GGA functional and with the Heyd-Scuseria-Ernzerhof (HSE [61]) hybrid functional are compared to experiment in Fig. 2. In the displayed frequency range theory and experiment show the presence of two closely spaced peaks in the case of the semiconducting (15,10) SWNT, and of only one peak in the case of the metallic (10,10) SWNT. The resonance peaks were attributed in ref. 33 to symmetry allowed transitions between valence and conduction bands with same band index, i.e. the $S_{33}$ and the $S_{44}$ transitions, respectively, in the (15,10) SWNT, and the $M_{11}$ transition in the (10,10) SWNT. In Fig. 2 the theoretical spectra were Lorentzian broadened with a width of 0.09 eV[34], and the intensities of the $S_{33}$ and of the $M_{11}$ features were scaled to set them equal to the corresponding experimental peaks. The $S_{44}$ intensity was not separately rescaled as the same scaling factor was used for the whole spectrum. As expected from previous calculations[53,54] the PBE peaks are red shifted by ~0.3–0.4 eV relative to experiment, while the HSE peaks are within 0.1 eV of experiment. It has been argued in ref. 34 that excitonic effects affect the RS features, particularly in the case of semiconducting nanotubes while screening significantly reduces those effects in metallic nanotubes. The adiabatic approximation for the exchange-correlation kernel does not account for excitonic effects in infinite periodic systems[48,62,63] and so the calculations reported in Fig. 2 cannot account for the fine details of the lineshape discussed in ref. 34. We remark, however, that in our study of detectability of nanoparticle growth under arc discharge conditions, we do not need an overly accurate description of the lineshape features but rather we need quantitative estimates of how the frequency-dependent polarizability of carbon nanoparticles changes with size and shape of the nanoparticles. We consider TDDFT in its simpler and less computationally expensive version provided by PBE to be sufficiently accurate for our purpose, and we use that approach in the calculations reported in the rest of the paper. The electronic and optical properties of infinite periodic carbon nanotubes have been widely discussed in the literature: see e.g. refs 64–66, the reviews in refs 67,68 and the references therein for a more complete discussion of how the various approximations, from empirical to first-principles, account for the optical properties of nanotubes.

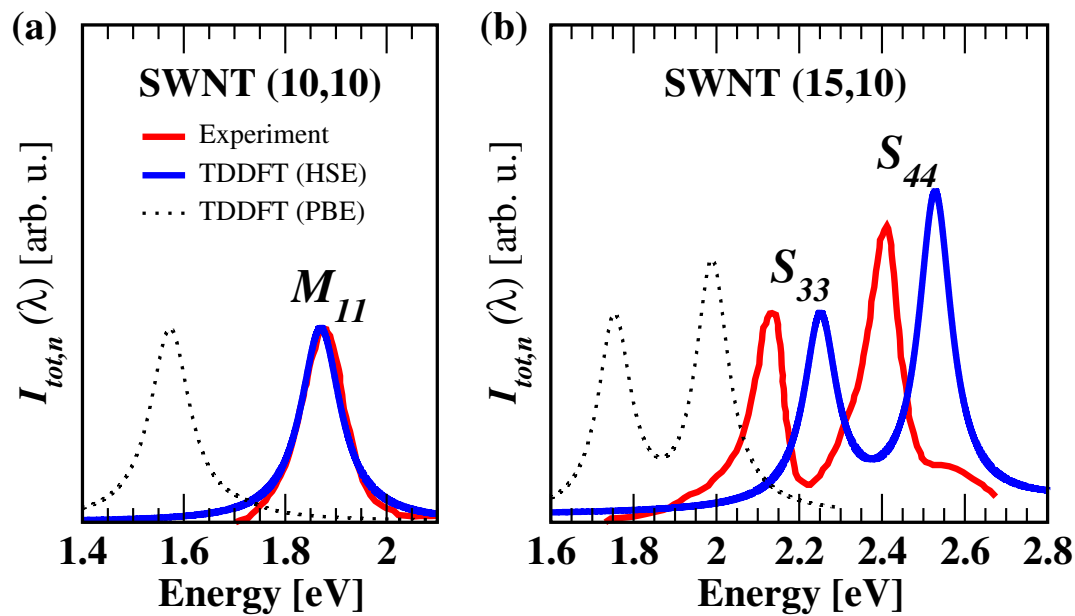

**Figure 2.** Theoretical and experimental RS spectra ($I_{tot,n}(\lambda)$) as a function of the incident photon energy for (**a**) the $M_{11}$ transition in the metallic (10,10) SWNT and (**b**) the $S_{33}$ and $S_{44}$ transitions in the semiconducting (15,10) SWNT. The theoretical spectra were computed with TDDFT using the Heyd-Scuseria-Ernzerhof (HSE[61]) and the Perdew-Burke-Ernzerhof (PBE[60]) exchange-correlation functionals. The experimental data were taken from ref. 33.

**Rayleigh Spectra in Arc Environment.** Carbon nanoparticles with different aspect ratios should be present in the arc discharge chamber in the early stage of SWNT growth. As nanotubes of given diameter grow in length their aspect ratio increases. We are thus interested in understanding how the RS spectrum changes for a particle of fixed diameter and evolving aspect ratio.

In the early stages of growth the gas-particle mixture in the arc chamber should contain asymmetric nanoparticles of small size in low concentration. In order to understand how the growing asymmetry of a nanoparticle affects the RS spectrum we considered a set of nanoparticles ranging from $C_{60}$ to a few short (5,5) CNTs capped on both sides by hemi $C_{60}$'s. All these nanotubes have the same diameter ($d$) of $C_{60}$, here the prototypical spherical nanoparticle, but have otherwise different lengths ($l$), as shown in Fig. 3(a). We consider, in particular, three capped CNTs with $l$ equal to 17.0, 24.5, and 32.0 Å and corresponding aspect ratios $\chi = l/d$ equal to 2.4, 3.5, and 4.5, respectively. Short nanotubes as the ones considered here are likely to be present in the early stages of synthesis. Indeed capped SWNTs as short as ~20 Å have been observed by TEM during CVD growth of CNTs on Fe nanoparticles[15].

With increasing aspect ratio the polarizability of the nanoparticles becomes more anisotropic, i.e. the longitudinal polarizability ($\alpha_\parallel$) becomes systematically larger than its transverse counterpart ($\alpha_\perp$), as shown in Fig. 3(b). The spectral range shown in the figure corresponds to the wavelength interval 390–1500 nm, which covers the visible spectrum and extends into the infrared region. The wavelengths of the lasers typically used for RS spectroscopy fall in the energy window of Fig. 3(b). In the low energy range (0.8–1.8 eV) each nanoparticle displays a characteristic resonance that moves systematically to lower energy with increasing aspect ratio. The resonance occurs at 3.38 eV (367 nm) in $C_{60}$ (not visible on the window of Fig. 3(b)) and moves to 1.39 eV (892 nm) in $C_{140}$, to 1.16 eV (1069 nm) in $C_{200}$ and to 1.02 eV (1215 nm) in $C_{260}$. It should be remarked that the states associated with this resonant transition have large weight on the two capping hemi $C_{60}$. The transition is characteristic of finite CNTs capped at both ends. It disappears in the infinite periodic (5,5) SWNT, which is metallic. In that case the lowest allowed transition occurs at 2.82 eV in our calculation.

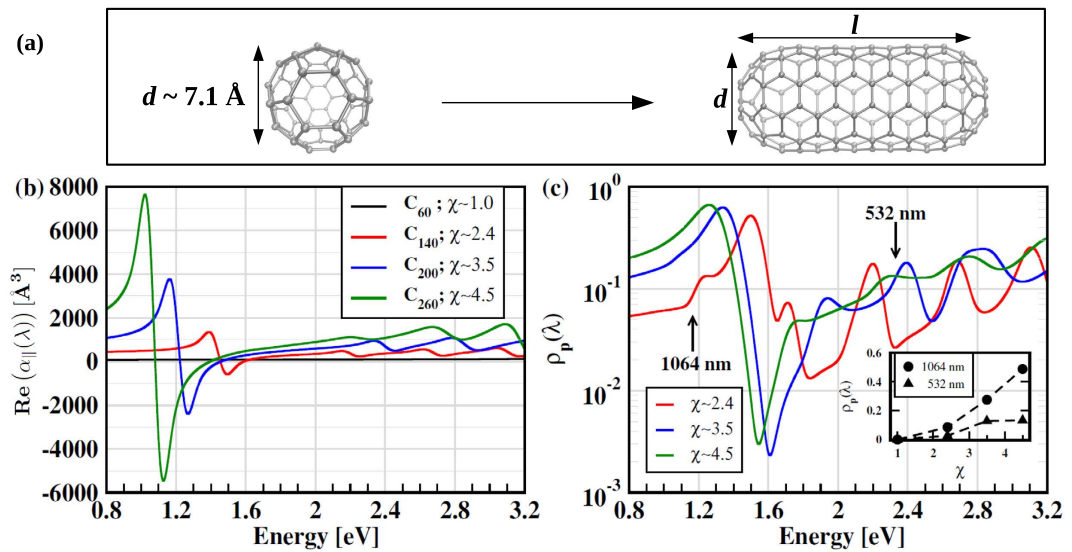

**Figure 3.** (**a**) Model nanoparticles. On the left, a symmetric $C_{60}$ nanoparticle is shown. On the right, an asymmetric bi-capped nanoparticle is shown. The capped-CNT has the same diameter ($d$) of the $C_{60}$ fullerene and length $l > d$. (**b**) Real part of the longitudinal polarizability ($\alpha_{\parallel}(\lambda)$) of nanoparticles with different aspect ratios ($\chi = l/d$) as a function of the incident photon energy. The polarizability was Gaussian broadened with a width of 0.08 eV. (**c**) Depolarization ratio ($\rho_p(\lambda)$) corresponding to the scattering from nanoparticles. The inset shows the depolarization ratio ($\rho_p(\lambda)$) at the two characteristic wavelengths (1064 nm and 532 nm) of the Nd:YAG laser as a function of the aspect ratio ($\chi$) of the nanoparticles.

Next, we estimate whether the RS signals of the above finite nanotubes could be detected in presence of the background gas in the arc chamber. We assume that the filling gas is Helium at T ~ 1500 K and p ~ 1 atm, a typical condition in the arc periphery where nanotube growth takes place[69–72]. The spherical symmetry of atomic He does not cause any change in the polarization of light after scattering. Thus, the He gas does not contribute to the horizontally polarized scattered intensity given in equation (2). According to equation (4) the total scattering intensity of the He gas can be written as:

$$I_{tot,b}(\lambda) = \frac{n_b V \pi^2 I_L}{\varepsilon_0^2 \lambda^4 r^2} \alpha_b^2,$$ (5)

where, $n_b$ is the concentration of He and $\alpha_b$ is the isotropic static polarizability of He ($\alpha_b \approx 0.205$ Å$^3$, see ref. 73). Within the energy range considered here the frequency dependence of the atomic polarizability of He is negligible.

The polarization of the scattered light can be different from the polarization of the incoming light only due to scattering from asymmetric particles. Thus one can detect the presence of asymmetric particles in a mixture of nanoparticles and He by measuring the fraction of scattered intensity with polarization different from that of the incident light. The depolarization ($\rho_p$) due to the nanoparticles is defined as the ratio of the intensity of the horizontally and vertically polarized scattered light:

$$\rho_p(\lambda) = \frac{I_{VH,n}(\lambda)}{I_{VV,n}(\lambda)} = \frac{3\gamma^2(\lambda)}{45\alpha^2(\lambda) + 4\gamma^2(\lambda)}.$$ (6)

Figure 3(c) reports the depolarization resulting from some asymmetric nanoparticles. The key finding is that the depolarization is noticeable even for a small asymmetry of the nanoparticles, i.e. when the aspect ratio is as small as 2.4. The maximum of $\rho_p(\lambda)$ shifts towards lower energy with increasing aspect ratio. The inset of Fig. 3(c) shows the trend of the depolarization ratio at the fundamental (1064 nm) and second harmonic (532 nm) wavelengths of the Nd:YAG laser. The increase of $\rho_p(\lambda)$ with increasing aspect ratio suggests that the depolarization could be detected with Nd:YAG laser. The increase of the depolarization with the aspect ratio reflects the increase of the polarizability with the asymmetry of the nanoparticles. However, when the wavelength of light is near a resonance the trend can be violated, namely less asymmetric particles could produce a larger depolarization than more asymmetric ones.

We now examine the dependence of the total RS cross section on the asymmetry of the nanoparticles in presence of the He gas. According to equation (5) the total scattering intensity from the He gas having density $n_b$ is a known function of the laser wavelength. Thus the ratio of the total scattering intensity of the nanoparticles and that of the background gas is given, at wavelength $\lambda$, by:

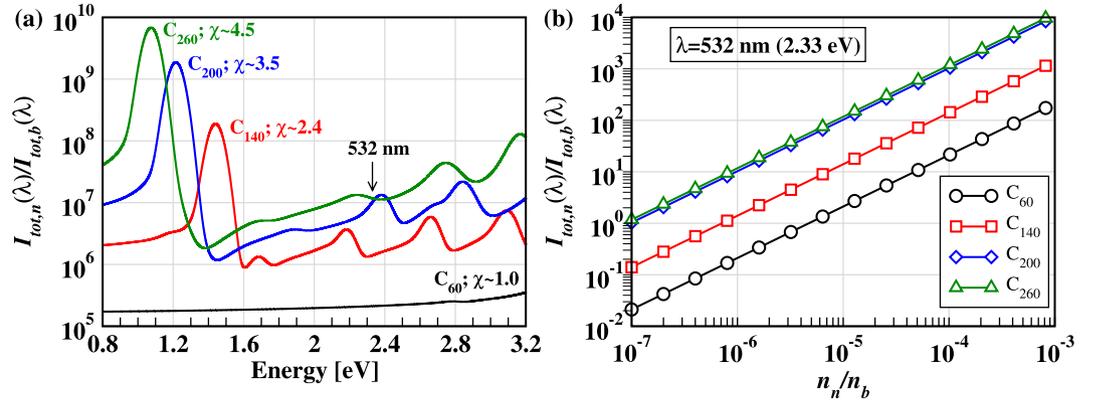

**Figure 4.** (**a**) The ratio of the total RS intensity of the nanoparticles ($I_{tot,n}(\lambda)$) and that of the background He gas ($I_{tot,b}(\lambda)$) (see equation 7) as a function of (**a**) the incident photon energy and (**b**) the partial concentration of the nanoparticles relative to the background He gas ($n_n/n_b$) at the second harmonic wavelength (532 nm) of the Nd:YAG laser. The RS intensity was Gaussian broadened with a width of 0.08 eV.

$$\frac{I_{tot,n}(\lambda)}{I_{tot,b}(\lambda)} = \left(\frac{n_n}{n_b}\right)\left(\frac{\alpha(\lambda)}{\alpha_b}\right)^2 \left(\frac{1 + \rho_p(\lambda)}{1 - \frac{4}{3}\rho_p(\lambda)}\right).$$

(7)

In equation (7) the relative RS intensity is proportional to the relative concentration of the nanoparticles, $n_n/n_b$, and depends quadratically on the relative polarizability of the nanoparticles, $\alpha(\lambda)/\alpha_b$. To examine the dependence of the relative RS intensity on the wavelength $\lambda$ we set ($n_n/n_b$) = 1 in equation (7) and report in Fig. 4(a) the relative RS intensity of different nanoparticles as a function of $\lambda$. While the relative RS intensity of $C_{60}$ is weak and almost flat over the entire energy window, the profiles corresponding to three asymmetric nanoparticles show large and distinct fluctuations that reflect the resonant behavior depicted in Fig. 3(b). This behavior is characteristic of each nanoparticle suggesting that it should be possible, at least in principle, to probe the early growth of specific nanotubes with a tunable laser.

The relative concentration of the nanoparticles with respect to the background gas is a major factor in determining the feasibility of *in situ* RS spectroscopy. In fact the RS signal of the nanoparticles will only be detected for relative concentrations higher than some threshold. We define this threshold as the relative concentration at which the RS intensity originating from the nanoparticles is equal to that produced by the background gas. The threshold depends on the specific nanoparticle and on the incident laser wavelength. As an example we report in Fig. 4(b) the variation of $I_{tot,n}(\lambda)/I_{tot,b}(\lambda)$ (equation (7)) with the relative concentration $n_n/n_b$ for $\lambda$ = 532 nm. The data reported in the figure refer only to four nanoparticles ($C_{60}$, $C_{140}$, $C_{200}$, and $C_{260}$) but are sufficient to give an idea of the qualitative trend. From Fig. 4(b) we see that for these nanoparticles the detectability threshold is in the range of relative concentrations between $10^{-7}$ and $10^{-5}$. Specifically, concentrations higher than $10^{-7}$ would be necessary to detect $C_{200}$ and $C_{260}$, while concentrations of the order of $10^{-5}$ and higher would be necessary for $C_{60}$. Larger carbon nanoparticles could be detected at significantly lower concentrations, as these nanoparticles will have generally larger RS cross-sections.

Finally, we consider the effect of the temperature on the RS spectra. Nanoparticle growth occurs in the periphery of the arc discharge where the temperature is estimated to be ~1500 K[69–72]. Nanoparticles in thermal equilibrium with the background gas have a Maxwellian distribution of velocities causing Doppler shifts in the scattered light. This in turn originates a symmetric Gaussian broadening of the RS signal. In the Knudsen regime, the normalized RS line shape centered at the mean frequency $\nu_0$ is given by[45]:

$$g_i(\theta, T, \lambda, m_i, \nu - \nu_0) = \frac{2}{\Delta\nu_T}\sqrt{\frac{\ln 2}{\pi}} \exp\left[-4\ln 2\left(\frac{\nu - \nu_0}{\Delta\nu_T}\right)^2\right],$$

(8)

where

$$\Delta\nu_T = \frac{|\mathbf{k}|}{2\pi}\sqrt{\frac{8k_B T \ln 2}{m_i}} \quad \text{and} \quad |\mathbf{k}| = \frac{4\pi}{\lambda}\sin\left(\frac{\theta}{2}\right).$$

(9)

In equations (8) and (9) $k_B$ is the Boltzmann constant, $m_i$ is the mass of the $i$-th particle, and $\theta$ is the scattering angle. In equation (8) $g_i(\theta, T, \lambda, m_i, \nu - \nu_0)$ integrated over $\nu$ is normalized to 1 and $\Delta\nu_T$ is the full width at half maximum. Then the thermally broadened RS signal of the $i$-th particle is:

$$S_i(\nu - \nu_0) \propto I_{tot,i}(\lambda)g_i(\theta, T, \lambda, m_i, \nu - \nu_0)$$

(10)

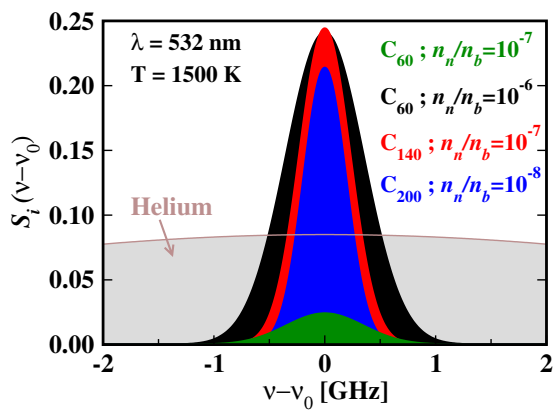

**Figure 5.** Doppler broadening of the RS spectral lines, $S_i(\nu - \nu_0)$ (see equations 8–10), corresponding to the background He gas and to nanoparticles with partial concentrations $n_n/n_b$ at a temperature of 1500 K. The mean frequency $\nu_0$ corresponds here to $\lambda = 532$ nm.

where, $I_{tot,i}(\lambda)$ is the total scattering intensity of particle $i$. The Doppler broadened spectra at $T = 1500$ K when $\lambda = 532$ nm are shown in Fig. 5 for the background He gas and three of the nanoparticles considered in this article. Due to the dependence of the full width at half maximum on the particle mass, the spectrum of the He gas has a very large broadening while the spectra of the nanoparticles become increasingly narrower with the mass of the nanoparticle. Because of the large difference in the relative broadenings, the RS signal of $C_{60}$ dominates over that of the background gas when the partial concentration of $C_{60}$ is of the order of $10^{-6}$ or higher, while the same happens for $C_{140}$ and $C_{200}$ at the even lower partial concentrations of $10^{-7}$ and $10^{-8}$, respectively. Thus Doppler broadening should enhance the sensitivity of RS spectroscopy by reducing the threshold concentration above which a given particle may be detected over the background signal. Moreover, at given temperature, laser wavelength, and scattering angle, the spectral width of the RS lines only depends on the mass of the scattering particle. This suggests that it should be possible, in principle, to identify the dominant masses of the nanoparticles that contribute to RS during nanoparticle growth, by measuring the line shape and its time evolution.

## Conclusions

Using TDDFT, we have calculated the frequency-dependent polarizability of carbon nanoparticles with different aspect ratios illustrating the increasingly more elongated nanoparticles that should be present in the early stage of CNT growth. From the frequency-dependent polarizabilities we extracted the corresponding RS spectra at typical thermodynamic conditions in the arc periphery region where growth occurs, including the effect of the background He gas. We found that even small asymmetries in the growing nanoparticles should be detectable by measuring the depolarization of the RS signal. Moreover, the presence of characteristic absorption resonances in the relevant optical range makes possible in principle to detect specific nanoparticles with tunable lasers.

Carbon deposits on a substrate placed in the growth region in the arc periphery suggest[74] the presence in that region of an average atomic concentration of carbon in excess of $10^{15}$ cm$^{-3}$. Under typical growth conditions a substantial fraction (~50% or more) of these atoms contribute to the SWNT yield[20,21,24]. If all these atoms formed small nanoparticles like $C_{60}$ in their early stage of growth the corresponding volumetric concentration of $C_{60}$ would be of the order of $10^{13}$–$10^{14}$ cm$^{-3}$, which is well above our estimated threshold for detectability by RS. Indeed our detectability threshold for $C_{60}$, $n_n/n_b$ greater than $10^{-6}$, corresponds to $n_n > 10^{12}$ cm$^{-3}$ for a typical concentration of the background He gas. Larger and/or more asymmetric nanoparticles would be detectable at lower concentrations, often substantially lower. These considerations suggest that RS spectroscopy is ideally suited for probing nanoparticles in the early stage of their formation under arc discharge conditions. Further improvement of the sensitivity should be possible by means of coherent Rayleigh-Brillouin scattering, which has been used in atomic gases[75] and could also be applied to characterize nanoparticles *in situ* under arc discharge conditions[42,76].

Real experimental conditions are more complex than the idealized situations considered here where only the RS response of a few selected nanoparticles has been considered. Ensembles of growing nanoparticles with different sizes, aspect ratios, combined with catalytic nanoparticles should be present under real experimental conditions, making highly nontrivial the disentanglement of the signals originating from the different nanoparticles. However, our study shows that different nanoparticles have characteristic detectable features, such as depolarization, resonant behavior, and mass-dependent line shape, which should help in the interpretation of the spectra, should the technique proposed here be used in real experiments.

Finally, while in this study we focused on the *in situ* characterization of nanoparticles under arc discharge synthesis, RS spectroscopy could also be useful to characterize nanoparticles during CVD synthesis.

## Methods

Ground-state and dynamic linear response electronic structure calculations were done with the QUANTUM ESPRESSO package[77]. In this package linear response theory within TDDFT is implemented in the frequency domain[78–80]. We adopted the semi-local approximation of Perdew-Burke-Ernzerhof (PBE[60]) for exchange and correlation in all calculations (static and dynamic). Calculations adopting the hybrid functional approximation of Heyd-Scuseria-Ernzerhof (HSE)[61] have also been performed for comparison in some cases. Only the valence electrons were treated explicitly and the interaction of the valence electrons with the nuclei and the frozen core electrons was modeled by pseudopotentials. In the PBE calculations we used ultrasoft pseudopotentials[81] from the QUANTUM ESPRESSO public library (http://www.quantum-espresso.org/pseudopotentials/; filename: "C.pbe-rrkjus.UPF")[82], requiring a plane wave kinetic (PW) energy cutoff of 30 Ry for the wave functions and of 180 Ry for the charge density (including the augmented density of the ultrasoft formalism). In the HSE calculations we used norm-conserving pseudopotentials[83] from the Qbox public library (http://fpmd.ucdavis.edu/potentials/; filename: "C_HSCV_PBE-1.0"), requiring a PW kinetic energy cutoff of 40 Ry for the wavefunctions and of 160 Ry for the charge density. To deal with finite nanoparticles we adopted a supercell approach, taking care that periodic images were separated by at least 15 Å of vacuum in each direction. Our tests indicate that the calculated dynamic polarizability is well converged with the above choice of parameters. The capped CNTs were constructed using the NanoCap package[84].

### Acknowledgements

This work was supported by U.S. Department of Energy, Office of Science, Basic Energy Sciences, Materials Sciences and Engineering Division, a Grant through DOE Contract No. DE-AC02–09CH11466. This research used resources of the National Energy Research Scientific Computing Center (NERSC), which is supported by the Office of Science of the U.S. Department of Energy. This research used resources of the Argonne Leadership Computing Facility at Argonne National Laboratory, which is supported by the Office of Science of the U.S. Department of Energy. Additional computational resources were provided by the Terascale Infrastructure for Groundbreaking Research in Science and Engineering (TIGRESS) High Performance Computing Center and Visualization Laboratory at Princeton University. B.S. is grateful to Hsin-Yu Ko, Alexandros Gerakis, and Predrag Krstić for helpful discussions. M.S. is grateful to Richard B. Miles for helpful discussions.

### Author Contributions

B.S., M.S. and R.C. conceived the project. B.S. performed the simulations. All authors analysed the results and reviewed the manuscript.

### Additional Information

**Competing financial interests:** The authors declare no competing financial interests.

**How to cite this article**: Santra, B. *et al. In situ* Characterization of Nanoparticles Using Rayleigh Scattering. *Sci. Rep.* **7**, 40230; doi: 10.1038/srep40230 (2017).

**Publisher's note:** Springer Nature remains neutral with regard to jurisdictional claims in published maps and institutional affiliations.